\documentclass[fleqn,11pt,twoside]{article}

\usepackage{amsthm,amsthm,amssymb, color, xcolor,epsfig, graphics, subfigure}

\usepackage{amsmath, graphicx, latexsym, lscape}

\makeatletter
\newcommand{\copyrightnote}[2]{{\renewcommand{\thefootnote}{}
 \footnotetext{\small\it
\begin{flushleft}
 \copyright \ #1   #2  
\end{flushleft}}}}

\newcommand{\Name}[1]{\begin{flushleft}
                       \LARGE \bf #1
                       \end{flushleft}\vspace{-3mm}}

\newcommand{\Author}[1]{\begin{flushleft}
                       \it #1 \end{flushleft}}

\newcommand{\Address}[1]{\begin{flushleft}
                       \it #1 \end{flushleft}}

\newcommand{\Date}[1]{\begin{flushleft}
                      \small  \it #1 \end{flushleft}}

\newcommand{\evenhead}{Author \ name}
\newcommand{\oddhead}{Article \ name}

\renewcommand{\@evenhead}{
\hspace*{-3pt}\raisebox{-15pt}[\headheight][0pt]{\vbox{\hbox to \textwidth
{\thepage \hfil \evenhead}\vskip4pt \hrule}}}
\renewcommand{\@oddhead}{
\hspace*{-3pt}\raisebox{-15pt}[\headheight][0pt]{\vbox{\hbox to \textwidth
{\oddhead \hfil \thepage}\vskip4pt\hrule}}}
\renewcommand{\@evenfoot}{}
\renewcommand{\@oddfoot}{}

\long\def\@makecaption#1#2{%
  \vskip\abovecaptionskip
  \sbox\@tempboxa{\small \textbf{#1.}\ \ #2}%
  \ifdim \wd\@tempboxa >\hsize
    {\small \textbf{#1.}\ \ #2}\par
  \else
    \global \@minipagefalse
    \hb@xt@\hsize{\hfil\box\@tempboxa\hfil}%
  \fi
  \vskip\belowcaptionskip}

\newcommand{\JNMPnumberwithin}[3][\arabic]{%
  \@ifundefined{c@#2}{\@nocounterr{#2}}{%
    \@ifundefined{c@#3}{\@nocnterr{#3}}{%
      \@addtoreset{#2}{#3}%
      \@xp\xdef\csname the#2\endcsname{%
        \@xp\@nx\csname the#3\endcsname .\@nx#1{#2}}}}%
}

\newcommand{\resetfootnoterule} {
  \renewcommand\footnoterule{%
  \kern-3\p@
  \hrule\@width.4\columnwidth
  \kern2.6\p@}
}

\renewcommand{\footnoterule}{}

\makeatother

\theoremstyle{definition}

\begin{document}

\renewcommand{\evenhead}{ {\LARGE\textcolor{blue!10!black!40!green}{{\sf \ \ \ ]ocnmp[}}}\strut\hfill S Carillo and C Schiebold}
\renewcommand{\oddhead}{ {\LARGE\textcolor{blue!10!black!40!green}{{\sf ]ocnmp[}}}\ \ \ \ \  Soliton equation: solutions via Bäcklund transformations}

\thispagestyle{empty}
\newcommand{\FistPageHead}[3]{
\begin{flushleft}
\raisebox{8mm}[0pt][0pt]
{\footnotesize \sf
\parbox{150mm}{{Open Communications in Nonlinear Mathematical Physics}\ \ \ \ {\LARGE\textcolor{blue!10!black!40!green}{]ocnmp[}}
\quad Special Issue 1, 2024\ \  pp
#2\hfill {\sc #3}}}\vspace{-13mm}
\end{flushleft}}

\FistPageHead{1}{\pageref{firstpage}--\pageref{lastpage}}{ \ \ }

\strut\hfill

\strut\hfill

\copyrightnote{The author(s). Distributed under a Creative Commons Attribution 4.0 International License}

\begin{center}
{  {\bf This article is part of an OCNMP Special Issue\\ 
\smallskip
in Memory of Professor Decio Levi}}
\end{center}

\smallskip

\Name{Soliton equations:  admitted solutions and invariances via 
B\"acklund transformations}

\Author{Sandra Carillo$^{\,1, 2}$ and Cornelia Schiebold$^{\,3}$}

\Address{$^{1}$ Dipartimento Scienze di Base e Applicate
    per l'Ingegneria \\  {``Sapienza''}   Universit\`a di Roma,  16, Via A. Scarpa, 00161 Rome, Italy \\[2mm]
$^{2}$ Gr. Roma1, IV - Mathematical Methods in NonLinear Physics\\ National Institute for Nuclear Physics (I.N.F.N.), Rome, Italy
\\[2mm]
$^{2}$Department of Engineering, Mathematics and Science Education (IMD),\\ Mid Sweden University, Sundsvall, Sweden}

\Date{Received November 1, 2023; Accepted January 15, 2024}

\setcounter{equation}{0}
\begin{center}
{\bf To Decio}
\end{center}

  \small{\it This short note is dedicated to the memory of a friend who was an inspiration for me further to be an important scientist in soliton investigation. I will not mention the many important results he obtained in his long carreer but I prefer to mention a {\it personal memory}   by S. Carillo:{\rm                                                                                                                                                                                                                                                                                                                                                                                                   ``Among the many occasions I was so lucky to share my time with Decio, I may recall that Decio introduced me to {\it who is doing what} in the field playing the {\it older brother} role with me.  Thus, the first NEEDS conference  I attended to in Balaruc, was really a key step in my career and, possibly, in my life.  My deep gratitude and memory to Decio and his kind and measured style will never be forgotten.''
}}

\begin{abstract}

\noindent 
A couple of  applications of B\"acklund transformations in the study of nonlinear  evolution equations is here given. Specifically, we are concerned about third order nonlinear  evolution equations. Our attention is focussed on 
one side, on proving a new invariance  admitted by a third order nonlinear  evolution equation and, on the other one, on the construction of solutions.
Indeed, via B\"acklund transformations,  a {\it B\"acklund chart}, connecting Abelian as well as non Abelian   equations 
can be constructed. The importance of such a net of links is twofold since it indicates invariances as well as allows to 
construct solutions admitted by the nonlinear  evolution equations it relates.  
The present study refers to third order nonlinear  evolution equations of KdV type. On the basis of the Abelian 
 wide B\"acklund chart which connects various different third order nonlinear  evolution equations an
 invariance admitted by the {\it Korteweg-deVries interacting soliton} (int.sol.KdV)  equation is obtained and a related new explicit solution is constructed.
 Then, the corresponding non-Abelian {\it B\"acklund chart}, shows how to 
  construct matrix solutions of the mKdV equations: some	recently obtained	 solutions are reconsidered.	
\end{abstract}

\label{firstpage}


\section{Introduction}
\label{introd}
The relevance of B\"acklund transformations \cite{Levi1, Levi2} in  the study of {\it soliton equations} is well known 
according to some of the most well known  books concerned about them \cite{Ablowitz, CalogeroDegasperis, 
RogersShadwick, RogersAmes, RogersSchief, Gu-book}.  In particular, the focus is on the Korteweg-de Vries (KdV) 
equation, one of the most studied third order nonlinear evolution equations, which,  both in the scalar case  \cite{BS1, 
apnum} as well as in the non-Abelian one,  \cite{Carillo:Schiebold:JMP2009,SIGMA2016, JMP2018, EECT2019}, 
turns out to be connected to many other third order nonlinear evolution equations. Specifically, a novel invariance admitted by the KdV interacting soliton equation, introduced by  Fuchssteiner \cite{Fuchssteiner1987}  is proved.  Furthermore,  
recent results  concerned about operator equations which represent the non-Abelian  counterpart of the 
{\it scalar} equations are reconsidered.    A wide net of B\"acklund transformations, we termed B\"acklund chart 
\cite{Carillo:Schiebold:JMP2009,SIGMA2016, JMP2018, EECT2019},
relates the KdV equation, the potential Korteweg-de Vries (pKdV), the  modified Korteweg-de Vries (mKdV),  the KdV 
eigenfunction (KdV eig.) to  the KdV singularity manifold equation (KdV sing.) \cite{Weiss}.  The constructed B\"acklund chart represents a key 
tool to show invariances enjoyed by the equations it connects as well as to construct solutions they admit.  
Third order nonlinear evolution equations  of KdV type are considered. Specifically, they are all connected via B\"acklund transformations. A {\it B\"acklund chart}, depicts the many links which connect  the  different nonlinear evolution equations under investigation. The  latest B\"acklund chart  is illustrated in \cite{EECT2019}; its  construction  is directly related to results in \cite{Fuchssteiner:Carillo:1989a, apnum} further developped. Generalisations, in the case of noncommutative  nonlinear evolution equations are comprised \cite{Carillo:Schiebold:JMP2009, Carillo:Schiebold:JMP2011, SIGMA2016, JMP2018} while a comparison between the two different cases Abelian and non-Abelian, respectively, is studied in \cite{EECT2019}.

\medskip
 The opening Section \ref{Benno's-eq} is focussed on the  KdV interacting soliton equation, 
as it was termed by  Fuchssteiner \cite{Fuchssteiner1987}: it is, now, proved to admit a non trivial 
invariance which seems to be new. As a consequence, via such an invariance, we construct a family of stationary 
solutions, again new,  admited  by the KdV interacting soliton equation. 

  The subsequent  Section \ref{matrix}, is devoted to  the non-Abelian B\"acklund chart, constructed in 
\cite{Carillo:Schiebold:JMP2009,SIGMA2016, JMP2018}, and, in particular,  the matrix  mKdV equation.
  Some solutions it admits are shown. 
 Specifically, a general theorem obtained in \cite{Carillo:Schiebold:JMP2011} is applied to derive solutions in the case of the 
$2\times 2$ matrix mKdV equation according to the results in  \cite{NODYCON, NODYCON2, NODYCON3}. 

In the closing Section  
\ref{perspectives} some perspectives and open problems are mentioned.  

\section{The KdV interacting soliton equation}\label{Benno's-eq}
This Section is devoted to the  {\it KdV interacting soliton}  equation,  denoted, for short, as int. sol. KdV, is introduced  by
 Fuchssteiner in \cite{Fuchssteiner1987}. This equation was then, in \cite{Fuchssteiner:Carillo:1989a}, connected to the KdV, mKdV and KdV sing. and, more recently, \cite{apnum, EECT2019} also to the KdV eigenfunction equation  \cite{boris90, 14}. 
 B\"acklund transformations as well known can be applied to reveal new symmetry properties, as well as to construct solutions of non-linear evolution equations they connect. Both these viewpoints are adopted in the present short note. Hence,  first of all  the definition of B\"acklund transformation is recalled. Then, the connections among third order non-linear evolution equations  are retrieved and, finally, an invariance, enjoyed by the int. sol.  KdV equation,   is readily constructed. Notably, such an invariance as well as the corresponding solution  seem  to be new. Indeed, the enjoyed invariance  allows to construct a non trivial solution of the int. sol.  KdV equation.
 
 According to \cite{Fuchssteiner:Carillo:1989a}, the int. sol. KdV equation \footnote{The int. sol. KdV  equation appears also in \cite{Faruk1} where third order nonlinear evolution equations and their linearizability are studied.} equation is connected via B\"acklund  
transformations to the  Korteweg deVries (KdV),  the modified Korteweg deVries (mKdV),  and the 
 {\it Korteweg deVries singuarity manifold}  (KdV sing.),  introduced by Weiss in  \cite{Weiss} via the {\it Painlev\`e test} of integrability.
\subsection{Invariance}
 
In this subsection   an invariance property  enjoyed by the  int.sol.KdV equation is proved. 
 It can be trivially checked to be {\it scaling invariant} since 
 on substitution of  $\alpha s, \forall \alpha \in \mathbb{C}$, to $s$  it remains unchanged.  
Remarkably, on application of results in  \cite{apnum, EECT2019},  the following    further nontrivial invariances can be proved.
\medskip

\noindent {\bf Proposition \ref{prop2b}\label{prop2b}} \\ \noindent  
{\it  The int.sol.KdV  equation $\displaystyle{s_t =  s_{xxx} - 3  {{s_x s_{xx}}\over s} + {3 \over 2 }{{s_x}^3 \over {s^2} } }$ is  invariant under the transformation 
\begin{equation}
\text{\rm I}: ~~~ \widehat s ={{a s( cD^{-1}( s) +d)d- c s ( aD^{-1}( s) +b)}\over{(c D^{-1}( s) +d)^2}},\quad a,b,c,d\in \mathbb{C}   
~ \text{s.t.} ~ad-bc\neq 0,  
\end{equation}
where $D^{-1}$ is chosen such that $D\circ D^{-1}$ is the identity.}%
\footnote{
Often one assumes that $s(x,t)$ belongs to the Schwartz space $S$ of {\it rapidly decreasing functions} 
for each fixed $t$. Here $S({\mathbb R}^n):=\{ f\in C^\infty({\mathbb R}^n) : \vert\!\vert f \vert\!
\vert_{\alpha,\beta} < \infty, \forall \alpha,\beta\in {\mathbb N}_0^n\}$, where 
$\vert\!\vert f \vert\!\vert_{\alpha,\beta}:= sup_{x\in{{\mathbb R}}^n} \left\vert x^\alpha D^\beta f(x)
\right\vert $, and  $D^\beta:=\partial^\beta /{\partial x}^\beta$; throughout this article $n=1$.
Then one may define $D^{-1}$ by
$ 
\displaystyle{D^{-1}:=\int_{-\infty}^x d\xi. }  
$ 
In calculations other choices may be useful.
}

The proof, according to \cite{apnum}, is based on the    invariance under the M\"obius group of 
transformations 
\begin{equation}
\text{M}:~~ \widehat\varphi={{a\varphi+b}\over{c\varphi+d}},\qquad a,b,c,d\in \mathbb{C} \qquad \text{such that} \quad ad-bc\neq 0.
\end{equation}
enjoyed by the KdV singularity manifold equation 
\begin{equation}
\varphi_t = \varphi_x \{ \varphi ; x\} , \quad \text{where} \ \ \{ \varphi ; x \} :=
 \left( { \varphi_{xx} \over \varphi_x} \right)_x -
{1 \over 2 }\left({ \varphi_{xx} \over \varphi_x} \right)^2.
\end{equation}
Combination of such an invariance with the connection between the KdV eigenfunction and
 the KdV singularity manifold 
equation allows to prove  the proposition. 
Indeed, let
\begin{equation}
\text{M}: \widehat \varphi={{a\varphi+b}\over{c\varphi+d}}~~,~~ \forall a,b,c,d\in \mathbb{C} \vert ~ad-bc\ne 0
\end{equation}
 then, the  B\"acklund chart  in Fig. 1, 
 \begin{figure}[h!]\label{BC}
\begin{eqnarray*}
\boxed{\varphi_t \ =\  \varphi_x  \{ \varphi ; x\}} ~\, {\buildrel {\rm B}\over{\text{\textendash\textendash\textendash\textendash\textendash\textendash}}}~\boxed{s^2 s_t = s^2 s_{xxx} - 3 s s_x s_{xx}+ {3 \over 2 }{s_x}^3}\\
\!\!\!\!\updownarrow~ \text{M} ~~~\qquad\qquad\qquad\qquad\qquad~~~~~\updownarrow~ \text{I} ~~~~~\qquad\qquad\\
\boxed{\hat\varphi_t \ =\  \hat \varphi_x  \{ \hat \varphi ; x\}} \,~ {\buildrel\widehat {\text{\rm B}}\over{\text{\textendash\textendash\textendash\textendash\textendash\textendash}}}~\boxed{\hat s^2 \hat s_t = \hat s^2 \hat s_{xxx} - 3 \hat s \hat s_x \hat s_{xx}+ {3 \over 2 }{ \hat s_x}^3}
\end{eqnarray*}
\caption{Induced invariance  B\"acklund chart.}
\label{fig int-sol-KdV-chart}
\end{figure}
\medskip\noindent
where the B\"acklund transformations ${\rm B}$ and $\widehat{\rm B}$ are, respectively:
\begin{equation*}
\text{\rm B}: ~~~~\displaystyle{s -  { \varphi_x}=0~ }~~~~\text{and}~~~~~~~~~~\widehat {\text{\rm B}}: ~~~~ \displaystyle{\hat s -  {\hat \varphi_x}=0~,}
\end{equation*}
shows how the invariance I is constructed. Indeed, 
such an invariance  follows via combination of the M\"obius   transformation M with the two B\"acklund transformations ${\rm B}$ and $\widehat{\rm B}$.
An application of the invariance ${\rm I}$  indicates how to construct solutions of the { KdV interacting soliton} equation.
\subsection{An admitted explicit solution}
 An  example of a new solution admitted be the { KdV interacting soliton} equation is readily obtained starting from its invariance proved  in the previous subsection. To construct it, it is easy to check
 that $s(x,t)=k, \forall k\in {\mathbb R}\backslash \{0\}$, represents a solution of the { KdV interacting soliton} equation
\begin{equation}
\displaystyle{s_t =  s_{xxx} - 3  {{s_x s_{xx}}\over s} + {3 \over 2 }{{s_x}^3 \over {s^2} } }~.
\end{equation}\label{w}
When, in the M\"obius group we let the parameters  be
\begin{equation}
a=d=0, \, b=c=1, 
\end{equation}
it follows, from the invariance $I$,  that a further solution of the { KdV interacting soliton} equation is represented by
\begin{equation}
\widehat s(x,t)= -{{1}\over {k x^2 }}~,  ~~ \forall k\in{\mathbb R}\backslash\{0\}~.
\end{equation}

Further solutions  can be obtained in the same way. Notably, the same way of reasoning allows to constrcut solutions also in the  non-Abelian corresponding case. 

\section{Non-Abelian case:  solutions of  mKdV equation}\label{matrix}
This section is devoted to non-Abelian equations. In \cite{Carillo:Schiebold:JMP2009, Carillo:Schiebold:JMP2011}, operator equations, which can be considered as non-Abelian counterparts of third order nonlinear  evolution equations of KdV type are studied and an extended B\"acklund chart is constructed \cite{JMP2018}. The comparison between the Abelian and the non-Abelian B\"acklund chart \cite{EECT2019} shows a richer structure in the non-commutative case.  In  particular, we consider the special case when the operator is finite dimensional so that it admits a matrix representation. Thus, the aim is to emphasise the importance of B\"acklund transformations also when  solutions admitted by non-Abelian soliton equations are looked for.
 Solutions admitted by the matrix  equations are a subject of interest in the literature. The study presented, based on previous results \cite{Carillo:Schiebold:JMP2009, Carillo:Schiebold:JMP2011} further developed in \cite{NODYCON, NODYCON2}, 
is consistent with multisoliton solutions of the matrix KdV equation obtained by Goncharenko \cite{Goncharenko},  via a generalisation of the Inverse Scattering Method.  Accordingly, Theorem 3 in \cite{Carillo:Schiebold:JMP2011}   represents a generalisation of
 Goncharenko's multisoliton solutions. Solutions of the matrix mKdV equation  obtained in \cite{NODYCON2} (motivated by \cite{NODYCON}), where the solution formula in the case of a $d \times d$-matrix equation is presented, are reconsidered.
 Note that this is a 
 particular case of the operator  formula obtained in \cite{Carillo:Schiebold:JMP2011}. 
For further matrix  solutions we refer to \cite{Chen, Goncharenko, LRB, Schiebold2009, Schiebold2018, Hamanaka1, Sakh}.

 As an example, some  $2 \times 2$-matrix solutions of the mKdV equation are recalled from \cite{NODYCON2}.
 They are contructed on application of the following theorem.

\medskip
\noindent {\bf Theorem \label{class}\ref{class}} (\cite{preprint}, see also \cite{NODYCON2}) \\ 
 {\it 	For $N \in \mathbb{N}$, let $k_1,\ldots,k_N$ be complex numbers such that $k_i+k_j \not= 0$ for all $i,j$, 
	and let $B_1,\ldots,B_N$ be arbitrary ${\sf d}\times {\sf d}$-matrices. 
	
	Define the $N{\sf d}\times N{\sf d}$-matrix function $L=L(x,t)$ as block matrix $L = (L_{ij})_{i,j=1}^N$
	with the ${\sf d}\times{\sf d}$-blocks
	\[   L_{ij} = \frac{\ell_i}{k_i+k_j} B_j ,   \]
		where $\ell_i = \ell_i(x,t) = \exp(k_i x + k_i^3 t)$.
		
	Then
	\[   V =  \begin{pmatrix} B_1 & B_2 & \ldots & B_N \end{pmatrix}
				\Big( I_{N\sf d} + L^2 \Big)^{-1} 
			\begin{pmatrix} \ell_1 I_{\sf d} \\ \vdots \\ \ell_N I_{\sf d} \end{pmatrix} 
	 \]
	is a solution of the matrix modified KdV equation 
	\begin{equation*}
V_t \ =\  V_{xxx}+3   \{V^2 ,V_{x}\} ~~,~~ \textstyle{where}~~ \{V^2 ,V_{x}\}:= V^2 V_{x}+V_{x}V^2 ~~~\textstyle{(anticommutator)}
\end{equation*}
with values in the ${\sf d}\times{\sf d}$-matrices
	on every domain $\Omega$ on which $\det(I_{N{\sf d}}+L^2) \not= 0$.} 
	
\begin{figure}[t]
\begin{center} \quad \\[-0.5cm]\label{example_4a}
	\includegraphics[scale=0.75]{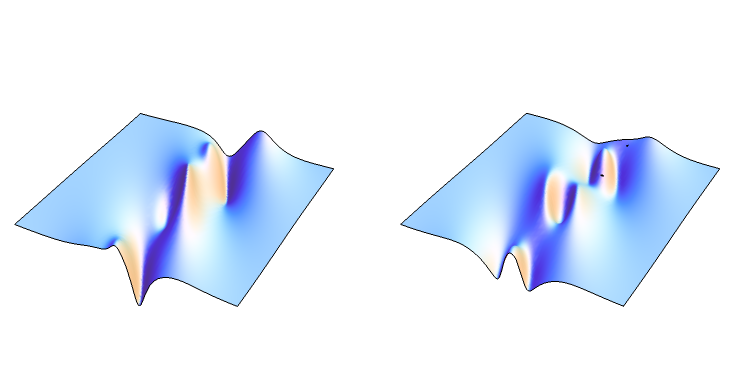} \\[-8ex]
	\includegraphics[scale=0.75]{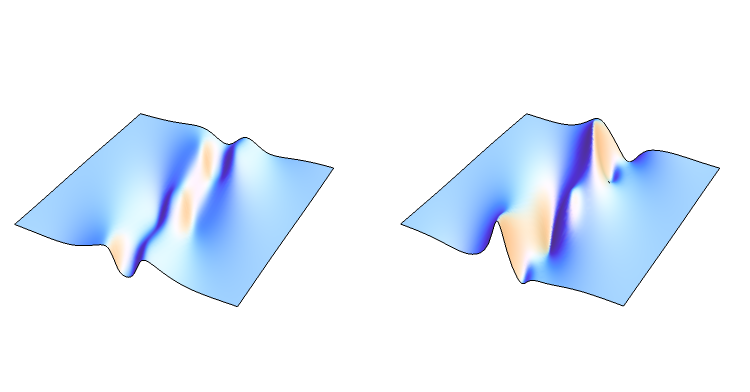} \\
	\caption	{ The solution  depicted in the case  ${\sf d}=2$, $k_1=1+i$, $k_2=\overline{k_1}=1-i$, 
	 $B_1 = \begin{pmatrix} i & -2 \\ 1+i & 2-i \end{pmatrix}$,	$B_2 =\overline{B_1}$,
        	 $-5\leq x \leq 5$ and $0\leq t \leq 2$, plot range   $(-3.5, 3.5)$.}	
\end{center}
\end{figure}

\begin{figure}[t] \label{example_4b}
\begin{center} \quad \\
	\includegraphics[scale=0.75]{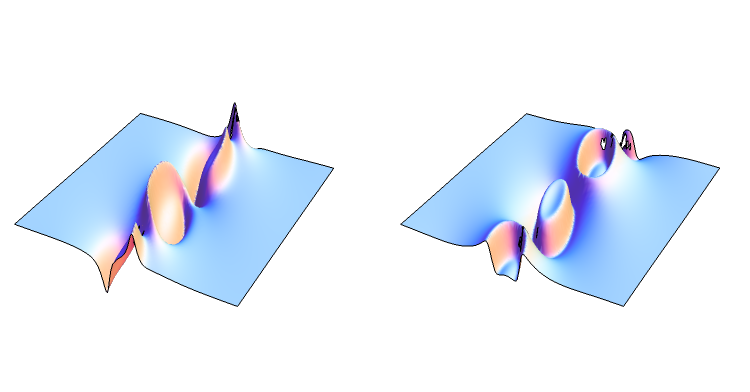} \\[-8ex]
	\includegraphics[scale=0.75]{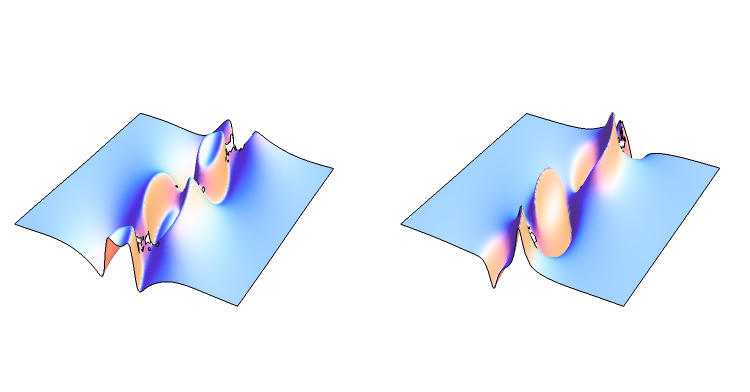} \\
	\caption	{ The solution  depicted in the case  ${\sf d}=2$, $k_1=1+i$, $k_2=\overline{k_1}=1-i$, 
	 $B_1 = \begin{pmatrix} i & -2i \\ 3i -1& -1 \end{pmatrix}$,	$B_2 =\overline{B_1}$,
    	 $-5\leq x \leq 5$ and $-1\leq t \leq 1$, plot range   $(-5.5, 5.5)$.}
\end{center}
\end{figure}

\medskip\noindent

Various kinds of solutions are obtained when complex parameters are considered.
According to what happens in the case of the scalar mKdV equation, where the input data $k$, $\overline{k}$, $b$, $\overline{b}$
produce a breather\footnote{
	Here $\overline{k}$ denotes the complex conjugate of $k$.}, 
 a bound state of a soliton and an antisoliton \cite{ZS},
also the matrix mKdV equation  admits breather solutions. 

\medskip\noindent

{\bf{Example 1} }
First examples are obtained when we set
 $k = 1 + {\rm i}$. For the corresponding scalar breather this implies velocity $=2$, and hence the plots are drawn for $(x+2t,t)$ giving a stationary picture.

To give an idea of some significant solutions admitted by matrix mKdV equation in the case ${\sf d}=2$, 
in Fig. 2 and Fig. 3 the solution is depicted
for the matrix parameters 
\begin{description}
	\item[a) {\sc Figure 2 
	}] \quad
		$B_1 = B=\begin{pmatrix} {\rm i} & -2 \\ 1+{\rm i} & 2 - {\rm i} \end{pmatrix},   B_2 = \overline{B} $, \\[0.1ex]
	\item[b) {\sc Figure 3 
	}] \quad
    		$B_1 =B=\begin{pmatrix} {\rm i} & -2{\rm i} \\ 3{\rm i}-1 & -1 \end{pmatrix}, B_2 = \overline{B} $.
\end{description}

\begin{figure}[t]\label{example_2_scalar}
\begin{center} \quad \\ [.5cm]
\includegraphics[scale=0.35]{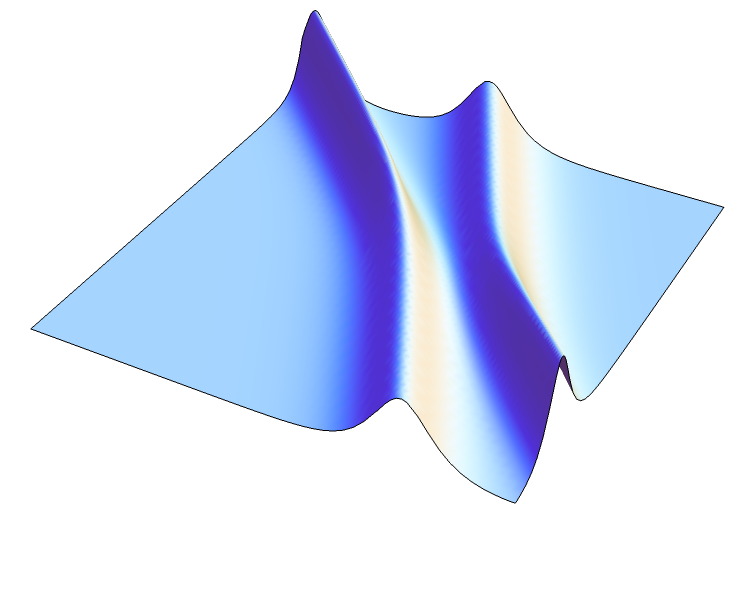} \\
\caption{${\sf d}=1$ with the input data $N=2$, $k_1=1$, $k_2=\sqrt{2}$, 
	and $b_1=b_2=1$ in Theorem \ref{class}}
	\end{center}
\end{figure}

\medskip\noindent

Next we turn to the 2-soliton case with real input data $k_1$, $k_2$, $b_1$, $b_2$.

\medskip\noindent

{\bf{Example 2} } 
Here we focus on the input data $N=2$, $k_1=1$, $k_2=\sqrt{2}$ in Theorem \ref{class}.

For comparison, first of all we depict, in Fig. 4, 
the \emph{scalar} 2-soliton (i. e. the case ${\sf d}=1$ in Theorem \ref{class}) generated with $b_1=b_2=1$.

Then we turn to the case ${\sf d}=2$,  where  in Fig. 5 and Fig. 6. the solutions are depicted 
generated with the matrix parameters 
\begin{description}
	\item[a) {\sc Figure 5 
	}] \quad $B_1 = \begin{pmatrix} 1 & 1 \\ 1 & 1 \end{pmatrix}$, 
						$B_2 = \begin{pmatrix}  \ 1 & -1 \\ -1 & \ 1  \end{pmatrix}$,
	\item[b) {\sc Figure 6 
	}] \quad $B_1 = \begin{pmatrix} 1 & 0 \\ 0 & 1 \end{pmatrix}$, 
						$B_2 = \begin{pmatrix} 0 & 1 \\ 1 & 0 \end{pmatrix}$.
\end{description}

\begin{figure}[h!] \label{3}\label{example_2b}
\begin{center} \quad \\[-1cm]
	\includegraphics[scale=0.75]{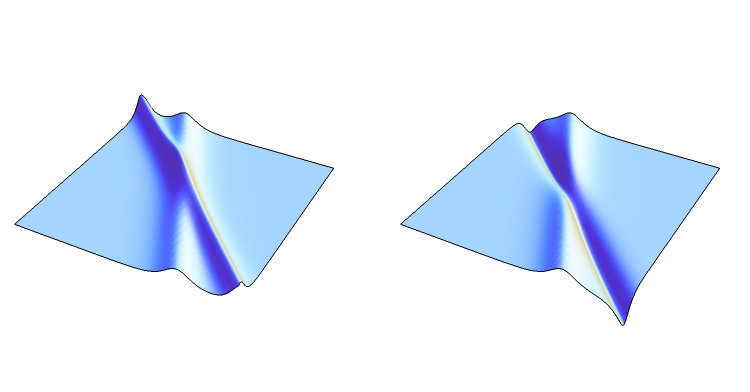} \\[-8ex]
	\includegraphics[scale=0.75]{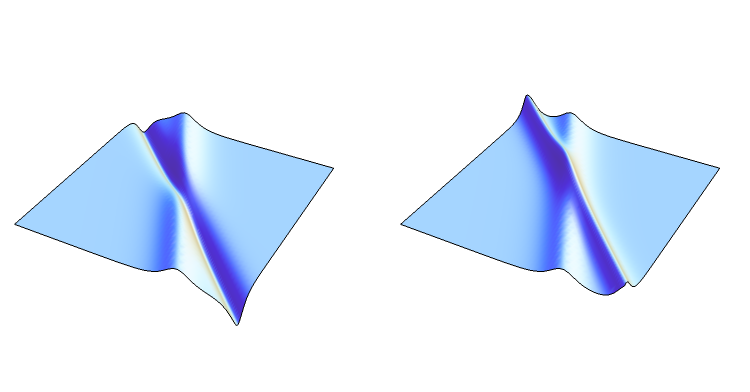} \\
		\caption{ The solution  depicted represents the case  ${\sf d}=2$, $k_1=1$, $k_2=\sqrt{2}$, 
	 $B_1 = \begin{pmatrix} 1 & 1 \\ 1 & 1 \end{pmatrix}$,	$B_2 = \begin{pmatrix} \ 1 & -1 \\ -1 & \ 1 \end{pmatrix}$, when
	$-10 \leq x \leq 10$ and $-5\leq t\leq 5$, plot range   $(-\sqrt{2}, \sqrt{2})$.}
\end{center}
\end{figure}

\begin{figure}[h!]\label{example_3}
{\begin{center} \quad \\ 
	\includegraphics[scale=0.75]{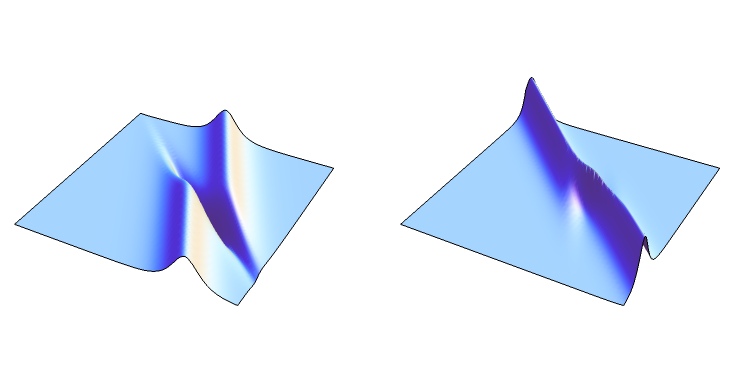} \\[-8ex]
	\includegraphics[scale=0.75]{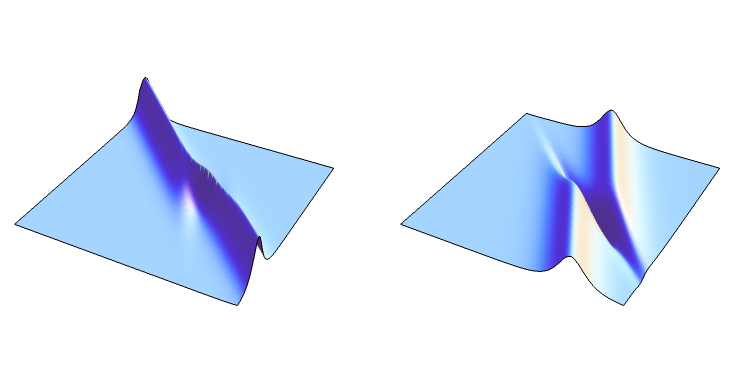} \\
	\caption{ The solution  depicted in the case  ${\sf d}=2$, $k_1=1$, $k_2=\sqrt{2}$, 
	 $B_1 = \begin{pmatrix} 1 & 0 \\ 0 & 1 \end{pmatrix}$,	\\ $B_2 = \begin{pmatrix}  0 & 1 \\ 1 &  0 \end{pmatrix}$, when
	$-10 \leq x \leq 10$ and $-5\leq t\leq 5$, plot range   $(-\sqrt{2}, \sqrt{2})$.}
\end{center}}
\end{figure}

\medskip\noindent

{\bf Comments and observations:}
\begin{itemize}
\item Obviously,  solutions which correspond to the  choice $k_1,\ldots, k_N \in \mathbb{R}$ 
and $B_1 = \ldots = B_N =: B$ (up to a common real multiple) where $B$ is real, are real-valued.
\item For $N=1$, Theorem  \ref{class} gives 
\begin{equation*}
	V =  \Big( I_d + \big( \frac{1}{2k} \ell B \big)^2 \Big)^{-1} \ell B ,
\end{equation*}
where $I_d$ denotes the $d$-dimensional identity matrix.

An example of a solution with a real spectral matrix $B$ which has complex Jordan form is given in \cite{NODYCON2}. Specifically, $B$ is the rotation by the angle $-\frac{\pi}{4}$,
\[   B= \begin{pmatrix} \ \ \frac{1}{\sqrt{2}} & \frac{1}{\sqrt{2}} \\ - \frac{1}{\sqrt{2}} & \frac{1}{\sqrt{2}} \end{pmatrix} . \]

\item The solutions depicted in Fig. 2 - Fig. 3 and Fig. 5 - Fig. 6 provide only some examples of the wide variety of solutions which are covered by Theorem \ref{class}.

\item In \cite{NODYCON3} first steps towards an asymptotic study of solutions from Theorem \ref{class} are given in the case $N=2$, see also \cite{Kielce}.
\end{itemize}

\section{Conclusions and perspectives}\label{perspectives} 
 
 To complete our work  we mention some of the themes which deserve to be further investigated.  In particular, as we 
already pointed out, the interest on B\"acklund transformations is twofold since they allow  to reveal new connections 
among equations, as well as they  indicate  a way to construct new solutions. A natural extension of the connections via 
B\"acklund transformations is represented by the extension to hierarchies. This aspect relies on the knowledge of the 
recursion operator admitted by at least one of the equations which appear in the B\"acklund  chart.  Then, according to 
\cite{Fuchssteiner:Carillo:1989a} also in the non-Abelian case  \cite{Carillo:Schiebold:JMP2009,SIGMA2016, JMP2018, 
EECT2019}, it follows that all the equations in the   B\"acklund chart admit a recursion operator: it can be obtained from 
the known recursion operator via B\"acklund transformations. Hence, the same B\"acklund chart follows to link the 
hierarchies generated by the recursion operator applied to the considered nonlinear evolution equations. Indeed, the algebraic 
properties which characterise  a hereditary recursion operator are preserved under B\"acklund transformations as well 
known \cite{Fuchssteiner1979, FokasFuchssteiner:1981}. Notably, the involved algebraic properties are preserved via 
B\"acklund transformations also when  non-Abelian nonlinear  evolution equations are studied \cite{Carillo:Schiebold:JMP2009}.

Indeed, promising perspectives as well as open problems can arise in the study of higher order nonlinear  evolution 
equations.

Notably, the B\"acklund chart  connecting 3rd oder (scalar) nonlinear evolution equations as well as the corresponding 
hierarchies \cite{Fuchssteiner:Carillo:1989a}, admit a non-Abelian counterpart. Such a non-Abelian B\"acklund chart is in 
\cite{Carillo:Schiebold:JMP2009, Carillo:Schiebold:JMP2011} with extensions  in \cite{SIGMA2016, JMP2018}.  
As testified by the study on  non-Abelian Burgers equation \cite{Ku,  GKT, Hamanaka2, Carillo:Schiebold:JNMP2012, 
MATCOM2017}, no matter which is the order of the nonlinear  evolution equations, the links established for the base members 
 naturally extend to the corresponding whole hierarchies. In particular, the non-Abelian Burgers B\"acklund chart exhibits a  structure which is richer than the 
corresponding Abelian one. 

A   B\"acklund chart \cite{Rogers:Carillo:1987b, BS1}, connects the Caudrey-Dodd-Gibbon-Sawata-Kotera and 
Kaup-Kuper\-shmidt hierarchies  \cite{CDG, SK, Kawa}.  All the involved equations are 5th order  nonlinear evolution equations; 
notably,  the B\"acklund chart linking them all shows an impressive resemblance to the one connecting KdV-type 
equations. Again, such B\"acklund chart can be extended to the corresponding whole hierarchies 
\cite{Rogers:Carillo:1987b, BS1}. Some preliminary results are given in \cite{CSF2023}.

\subsection*{Acknowledgements}
The financial support  of Gr. Roma1, IV - Mathematical Methods in NonLinear Physics National Institute for Nuclear Physics (I.N.F.N.), Rome, Italy, of \textsc{Sapienza}  University of Rome, Italy  and of the National Mathematical Physics Group (G.N.F.M.) - I.N.d.A.M., is gratefully acknowledged. 

\label{lastpage}
\end{document}